\documentclass[conference]{IEEEtran}
\IEEEoverridecommandlockouts
\usepackage{cite}
\usepackage{amsmath,amssymb,amsfonts}
\usepackage{algorithmic}
\usepackage{graphicx}
\usepackage{textcomp}
\def\BibTeX{{\rm B\kern-.05em{\sc i\kern-.025em b}\kern-.08em
    T\kern-.1667em\lower.7ex\hbox{E}\kern-.125emX}}

\usepackage{booktabs} 

\usepackage{comment}
\usepackage{subfig}
\usepackage{multirow}
\usepackage{url}
\usepackage{booktabs}
\usepackage{xspace}
\newcommand{\MATLAB}{\textsc{Matlab}\xspace}

\begin{document}

\title{Optimal Checkpointing for Secure Intermittently-Powered IoT Devices\\
\thanks{This research was supported in part by NSF Grants \#1553419 and
\#1319841. Any views expressed  are the authors' own and do not
necessarily reflect the views of the NSF.}
}

\author{\IEEEauthorblockN{Zahra Ghodsi}
\IEEEauthorblockA{
\textit{New York University}\\
New York, USA \\
ghodsi@nyu.edu}
\and
\IEEEauthorblockN{Siddharth Garg}
\IEEEauthorblockA{
\textit{New York University}\\
New York, USA \\
sg175@nyu.edu}
\and
\IEEEauthorblockN{Ramesh Karri}
\IEEEauthorblockA{
\textit{New York University}\\
New York, USA \\
rkarri@nyu.edu}
}

\maketitle

\begin{abstract}
Energy harvesting is a promising solution to power Internet of Things (IoT) devices. 
Due to the intermittent nature of these energy sources, one cannot guarantee forward progress of program execution.
Prior work has advocated for checkpointing the intermediate state to off-chip non-volatile memory (NVM). Encrypting checkpoints addresses the 
security concern, but significantly increases the checkpointing overheads. In this paper, we propose a new online checkpointing policy that \emph{judiciously} determines when to checkpoint so as to minimize
application time to completion while 
guaranteeing security. Compared to state-of-the-art checkpointing schemes that do not account for the overheads of encrypted checkpoints we improve execution time up to $1.4\times$.

\end{abstract}


\section{Introduction}\label{intro}

The Internet of Things (IoT) is envisioned as a network of devices that operate anytime from anywhere. IoT devices are expected to be small and guarantee perpetual and autonomous operation even in hard
to reach places. 
Energy harvesting has been proposed as the source of 
energy for such IoT devices \cite{liu:dac,eh}. 
Ambient energy can be harvested from many sources, including solar, wind, thermal, WiFi, radio frequency, biological and chemical~\cite{h2,h1,  h3}. 
The amount of energy harvested is highly time varying and the energy supply is intermittent. For this reason, processors with off-chip non-volatile main memory (and potentially even on-chip caches and register files) 
have been proposed \cite{ liu:dac,ma:hpca} to guarantee 
\emph{forward progress} of program execution. In these approaches, the processor saves its intermediate results in non-volatile memory when the energy supply is too low for the processor to operate.

Security is another critical concern with IoT devices. 
An attacker with physical access to an IoT 
device might be able to read out potentially sensitive data from the non-volatile memory~\cite{nvmat}.
Prior work secures data in IoT devices using non-volatile main memory, i.e. by   
encrypting and decrypting every write to and read from the non-volatile memory~\cite{Chhabra:i-nvmm,Enck:securingnvmm}. However, simply encrypting main memory is not sufficient. An attacker might capture an IoT device 
in the middle of program execution and recover a checkpoint of architectural state. In fact, prior work has shown that unencrypted intermediate
program state is an even greater security threat~\cite{Halderman}. Thus, we propose to encrypt the intermediate checkpoints of program state to guarantee confidentiality at all times during program execution.

\newcommand{\myt}{
\begin{tabular}{ |c|c|c| } 
 \hline
 Cipher & CP Energy (nJ) & CP Latency (cycles) \\ \hline
 None & 5.75 & 82 \\ 
 Prince & 66.90 & 164 \\ 
 AES & 410.45 & 164 \\
 \hline
\end{tabular}
}

\begin{figure}
\centering
\subfloat[][]{%
  \includegraphics[width=0.48\textwidth]{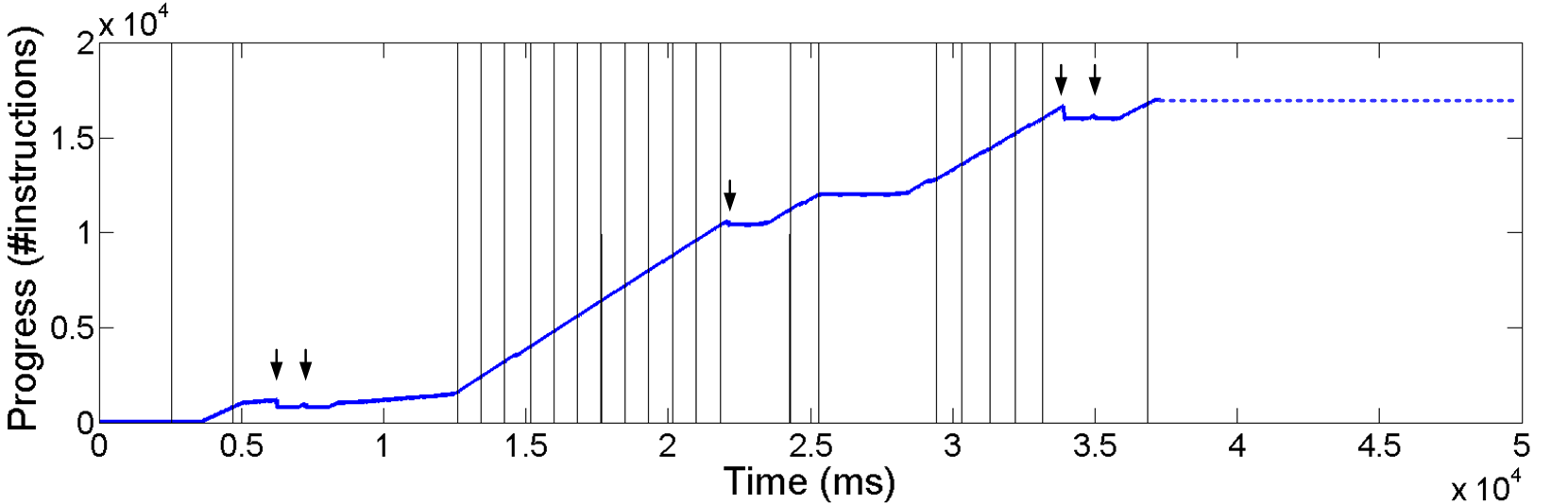}%
  \label{fig:motive1}
  }\par
\subfloat[][]{%
  \includegraphics[width=0.48\textwidth]{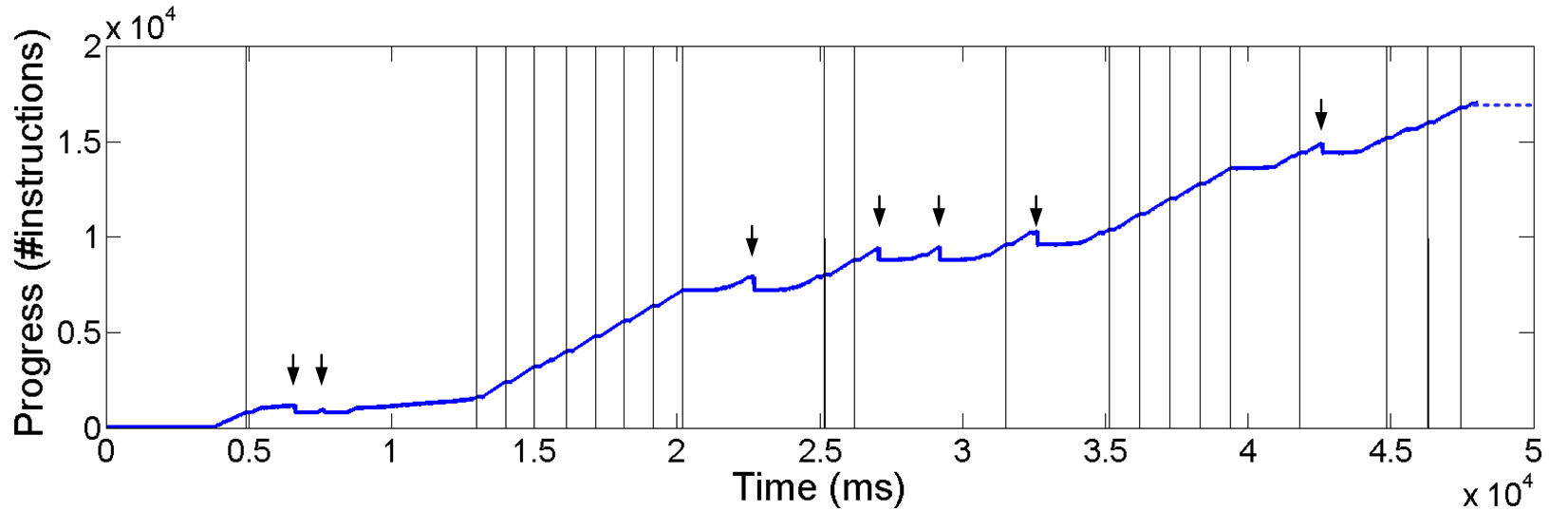}%
  \label{fig:motive2}
  }
\caption{Runtime execution progress of the FFT program running on harvested energy with (a) no encryption and (b) PRINCE~\cite{prince} as the block cipher. Vertical lines represent checkpoints and arrows show roll-backs due to energy failure. Encryption increases the average energy and latency of each checkpoint. Due to this overhead, the total number of roll-backs, time spent for each checkpoint and therefore the program execution time increases.}
\label{fig:motive}
\end{figure}

\begin{figure*}[t]
\centering
\includegraphics[width=1.0\textwidth]{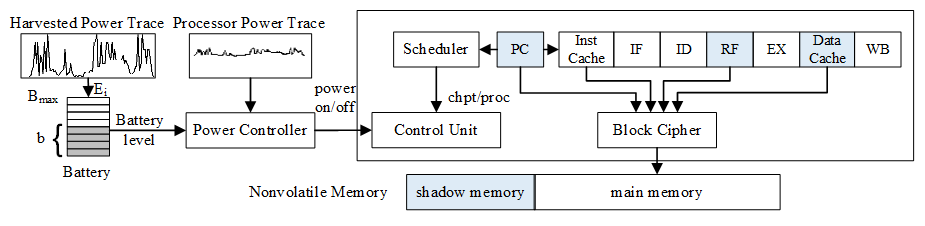}
\caption{Optimal checkpointing of secure IoT processors: System model with a finite-sized battery with capacity $B_{max}$ and random energy arrivals. At a checkpoint, the volatile IoT system state (Program Counter (PC), Register File (RF) and dirty cache lines) is encrypted and saved into NVM.}
\label{fig:model}
\end{figure*}

Encryption increases the energy and latency overhead of checkpoints, by $11.6\%$ and $200\%$ respectively for PRINCE~\cite{prince} block cipher. 
Finding the optimum checkpoint placement policy when the energy source is intermittent is crucial, and becomes even more significant when checkpoints have to be encrypted for security. Naively replacing unencrypted 
checkpoints (Fig~\ref{fig:motive1}) 
with encrypted checkpoints (Fig~\ref{fig:motive2}) 
can significantly 
increase the number of roll-backs and total program execution time. 
This motivates the need for a 
smart checkpointing policy that 
places checkpoints judiciously.
The checkpointing policy must
account for multiple 
factors including
the energy level of the battery, the stochastic behaviour of the  
harvested energy (which differ from one source to another), when 
the previous checkpoint was taken, and the 
expected time to program completion. Prior checkpointing policies account for some of these effects, 
for instance, periodic checkpointing~\cite{liu2016lightweight}
or checkpointing when the available energy in the battery falls below a threshold~\cite{hibernus,quick}.

We formalize the online checkpointing problem as a 
Markov decision process, and compute the optimal checkpointing 
policy offline using Q-learning \cite{Watkins1992}.
Online decisions are made using a table including the optimum action which is obtained from the Q-table (stored in memory).
Our solution: (i) is the first approach that simultaneously
(and explicitly) accounts for: the 
current battery level, previous time a checkpoint was taken, and 
the program's forward progress in informing checkpointing decisions;
(ii) accounts for the stochastic nature of the energy source,  
checkpointing overheads, and processor energy consumption; 
(iii) uses a model-free approach to solving the Markov decision process that can be trained using empirically obtained and synthetically generated traces of harvested energy;
and (iv) offers computational and energy efficient hardware solutions to make online decisions.


The rest of the paper is organized as follows: Section~\ref{related} 
reviews the recent work in this area, while Section~\ref{method} 
explains system model and the proposed Q-learning based 
online checkpointing policy. 
Section~\ref{experiments} describes our experimental setup, results and 
compares the proposed approach to prior art. We conclude in Section~\ref{conclusion}
to future work.

\subsection{Related Work}\label{related}


Checkpointing has been used for fault tolerance
in computing systems~\cite{cp1,cp2}, allowing the 
processor to roll back to the last valid checkpoint and recover the processor 
state in case of failure. 
Okamura et al.~\cite{okamura2004dynamic} propose a dynamic checkpointing scheme based 
on Q-learning for fault-tolerance, 
but assume random failures unrelated to energy harvesting or power failures.

Checkpointing has also been used to guarantee forward progress
in intermittently powered 
devices \cite{hibernus,quick,liu2016lightweight,ma:hpca,tpc,mementos}. 
Mementos \cite{mementos} 
inserts trigger points in the software
at compile time and, at run time, checkpoints when 
the stored energy level falls below a threshold at a trigger point. 
QuickRecall~\cite{quick} 
and Hibernus~\cite{hibernus} propose a policy based on two thresholds, one that triggers checkpoints 
(when the energy level falls below a \emph{low} threshold)
and one that determines when to start re-execution (when the energy level rises 
above a \emph{high} threshold).
A similar approach is used in \cite{ma:hpca}, except that 
they assume access to a processor with on-chip non-volatile state.
Hardware support for checkpointing that uses two counters for number of instructions and number of stores is proposed in~\cite{liu2016lightweight}. A checkpoint is performed if either of the two counters exceeds its threshold.

From a security stand-point, several proposals exist on main memory encryption for security, \cite{Chhabra:i-nvmm,Enck:securingnvmm} but they do not
consider battery-operated intermittently powered processors. Finally, Q-learning has been used in other online decision making contexts such as communications and power management \cite{1581377,790549,Tan:2009}.


\section{Online Checkpointing}\label{method}
We describe an online checkpointing framework for secure IoT processors starting with the system model, a mathematical formulation 
of the optimal checkpointing problem, and a learning based solution.

\subsection{System Model}
Our target IoT system shown in Fig~\ref{fig:model},
includes an in-order processor with conventional volatile caches and registers running on harvested energy and an off-chip non-volatile main memory. Although the proposed techniques are agnostic to the NVM technology, in Section~\ref{experiments} we empirically evaluate an RRAM-based NVM. For securing the checkpoints, our system encrypts main memory similar to prior work~\cite{Chhabra:i-nvmm,Enck:securingnvmm}. 
Data blocks written to and read from main memory are encrypted and decrypted, respectively, 
using a light-weight block cipher PRINCE~\cite{prince}.

The system is powered by harvested energy which is highly intermittent. To smooth out large temporal variations in harvested energy, the IoT device has on-board energy storage\footnote{On-board energy storage can range from a simple super-capacitor to a  
battery management IC with an in-built battery tailored for ultra-low power energy harvesting devices such as  the TI BQ25504~\cite{TIBQ}.}. To guarantee forward progress of program execution, the processor state is checkpointed in NVM, and to guarantee security, all checkpoints are encrypted before being written to the NVM.
Shown as shaded in Fig \ref{fig:model}, a checkpointed IoT system state consists of: 
(i) the program counter (PC), 
(ii) the contents of the register file (RF);
and (iii) all dirty cache lines in the L1 data cache\footnote{We assume a single-level cache hierarchy.}. 
The NVM has a \emph{shadow memory} to store  a checkpoint. 

When rolling back to the last checkpoint, the state of the system consisting of volatile and non-volatile states should be maintained consistently. The volatile states are stored and recovered from the shadow memory, therefore they remain consistent. Ransford et. al.~\cite{ransford2014nonvolatile} showed that checkpointing and recovering the program state can lead to inconsistency in the NVM. All writes to NVM that happen after a checkpoint change the non-volatile state of the program. In case of an energy failure, the processor rolls back to the checkpointed state and the volatile states are recovered correctly, however, the non-volatile state has changed since the checkpoint and the NVM will be inconsistent.

To address this issue, previous work has proposed enforcing a checkpoint between every non-idempotent memory access pairs~\cite{xie2015fixing}, or versioning the inconsistent data~\cite{ransford:atomic} so that every non-idempotent memory access pair is eliminated. Liu et. al.~\cite{liu2016lightweight} propose a hardware solution in which all stores are regarded as speculative and are delayed. These stores are written back to non-volatile memory only when the program reaches the next checkpoint. To keep the overhead of processing low and avoid placing an excessive number of checkpoints, we simply expand the shadow memory to keep track of the memory locations which are written to between two checkpoints. For each write to the non-volatile memory, the previous value in the memory is copied to the shadow memory. If the next checkpoint is performed successfully, these values are discarded and otherwise they are restored along with the saved volatile state.


Finally, the IoT processor is provisioned with the scheduler module (Fig~\ref{fig:model}), 
that determines \emph{when} checkpoints are performed. The design and  evaluation of a security-aware checkpoint scheduler is the primary contribution of this work. 



\begin{figure}
\centering
\includegraphics[width=0.45\textwidth]{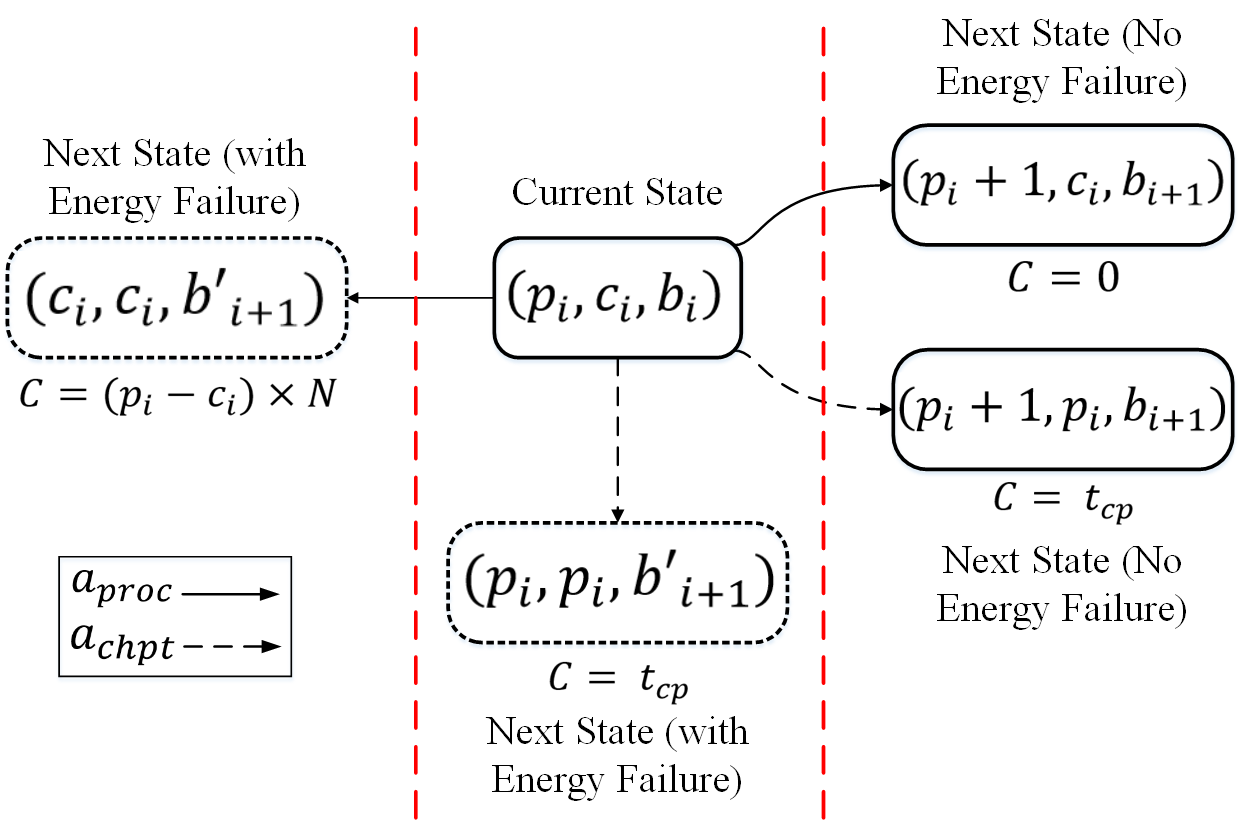}
\caption{State transition diagram for actions $a_{chpt}$ (checkpoint) and $a_{proc}$ (no checkpoint) with and without an energy failure event. Each state is defined as $s_{i}= (p_i, c_i, b_i)$ where $p_i$ represents the progress made so far, 
$c_i$ is where the last checkpoint was performed and $b_i$ is the current battery level. For each action, $C$ indicates the immediate cost incurred from that action.}
\label{fig:mdp}
\end{figure}

\subsection{Problem Formulation}

The checkpoint scheduler 
decides whether to checkpoint or not every $N$ 
instructions. Every block of $N$ executed instructions is 
referred to as an \emph{interval}.
To track forward progress, the 
scheduler maintains two 
run-time counters:
(i) 
a \textit{Progress Counter (PrC)} which counts the number of intervals of actual
forward progress that the program has made; and 
(ii) a \textit{Checkpoint Counter (CC)}
which records when the last checkpoint was taken. Both
counters are initialized to zero when the 
program begins.
PrC is incremented every time $N$ 
instructions execute successfully. 
If a checkpoint is made, CC is updated to the current PrC. Finally,  
on energy failure, the processor rolls back to the last checkpointed state and this resets PrC to the current value of CC. 


We formulate the execution of a 
program with online checkpointing 
as an instance of a
Markov decision process (MDP) consisting of
 a set of states $S$, a set of actions $A$, a stochastic transition function $T:S\times A \times S \rightarrow [0,1]$ which gives the probability of moving from one state to another for each action, and an immediate cost function $C:S \times A \times S \rightarrow \mathbb{R}$.

In our formulation, the state of the Markov decision process is given by the current values of the progress counter (PrC), the checkpoint counter (CC) and the battery level. Specifically, the state at the end of each control interval $i$
is given by 
$s_{i}= (p_i, c_i, b_i)$ 
where $p_i$ is the PrC value (representing the progress made so far), 
$c_i$ is CC (when the last checkpoint was performed) and $b_i$ is the current battery level. 
The set of actions is $A \in \{a_{chpt},a_{proc}\}$, where the action $a_{chpt}$ represents taking a checkpoint while the action $a_{proc}$ represents 
proceeding without a checkpoint. 

If the scheduler decides to checkpoint, a cost $t_{cp}$
corresponding to the latency of checkpointing is incurred. Assuming no energy failure during the next interval, the next state is $s_{i+1} = \{p_{i}+1, p_{i}, b_{i+1}\}$. On the other hand, if there is an energy failure in the next interval, the next 
state is $s_{i+1} = \{p_{i}, p_{i}, b'_{i+1}\}$. In both cases, $b_{i+1}$ and $b'_{i+1}$ are the 
new battery levels accounting for the energy cost of checkpointing, the net energy received/consumed in interval $i+1$.  

On the other hand, if the scheduler decides \emph{not} to checkpoint, there is no immediate cost. Assuming no energy failure during the next interval, the next 
state is $s_{i+1} = \{p_{i}+1, c_{i}, b_{i+1}\}$.
However, if there is an energy failure in the next interval, the next state is $s_{i+1} = \{c_{i}, c_{i},  b'_{i+1}\}$, incurring a cost of $(p_i-c_i) \times N$ representing the lost progress. The state transition diagram is shown in Fig~\ref{fig:mdp}.

The optimal policy balances the cost of checkpointing with the cost of recovering from an energy failure by rolling back to the last checkpoint. The total cost can be defined as the time overhead of checkpointing plus the re-execution time due to roll-backs. Given the Markov decision process specification, our
goal is to find an optimal policy 
$\pi^*:S \rightarrow A$ which
minimizes the total  expected cost as defined above, i.e., the expected overhead of the checkpointing policy relative to the baseline uninterrupted execution time.
The optimal policy can be obtained by solving for a fixed-point of the  Markov decision process
given the stochastic model for the harvested energy and system statistics i.e.,  the probability of energy failure in future control intervals and a distribution of next state battery levels. However, this model would be hard to obtain and can be highly inaccurate. 
Hence, the proposed system learns an optimal policy using experimentally obtained traces of harvested energy and system statistics.


\subsection{Offline Learning of Checkpointing Policy}\label{model}


The system uses the Q-learning algorithm to find the optimal policy. This algorithm assigns a 
Q-value to each state-action pair 
$Q: S \times A \rightarrow \mathbb{R}$ that, once the algorithm converges, represents the expected total cost of executing action $a$ in state $s$ and choosing greedy actions afterwards. The algorithm starts by 
initializing all Q-values arbitrarily, and iteratively updates by simulating the system.  In iteration $i$,  
the algorithm chooses the action $a$ that results in the smallest Q-value for the  current state, simulates the system,  and observes the next state and corresponding cost. It then updates the Q-values
as follows:
\begin{equation}
\begin{aligned}
Q(s_i,a) & \leftarrow Q_i(s_i,a) + \\ 
& \alpha ( C(s_{i},a) + \gamma \min_{a' \in A} Q_i(s_{i+1},a') - Q_i(s_{i},a)).
\end{aligned}
\label{eq:q}
\end{equation}
Here, 
$C(s,a)$ is the checkpointing cost $t_{cp}$ 
for checkpointing and
roll-back cost $(p_{i} - c_{i})\times N$ in the event of an energy  failure for no checkpointing. 
The parameter
$\gamma$ discounts future costs, 
and $\alpha$ determines the learning rate, i.e., 
how strongly the Q-values are overwritten by new ones after each iteration. With certain constraints on the learning rate \cite{Tsitsiklis1994}, the 
Q-values have been proven to converge to 
those corresponding to the optimal policy $\pi^{*}$. Empirically, we found that 
using a variable learning rate worked best in practice. 
\begin{equation}
\alpha_{(s,a)} = \frac{1}{n(s,a)}
\end{equation}
where $n(s,a)$ is the number of times a state-action pair $(s,a)$ is visited.
Further, we use the $\epsilon-greedy$ algorithm which chooses an 
action at random with probability $\epsilon$ and follows the greedy strategy with probability $1-\epsilon$ where $0<\epsilon<1$. 


\begin{figure}
\centering
\includegraphics[width=0.45\textwidth]{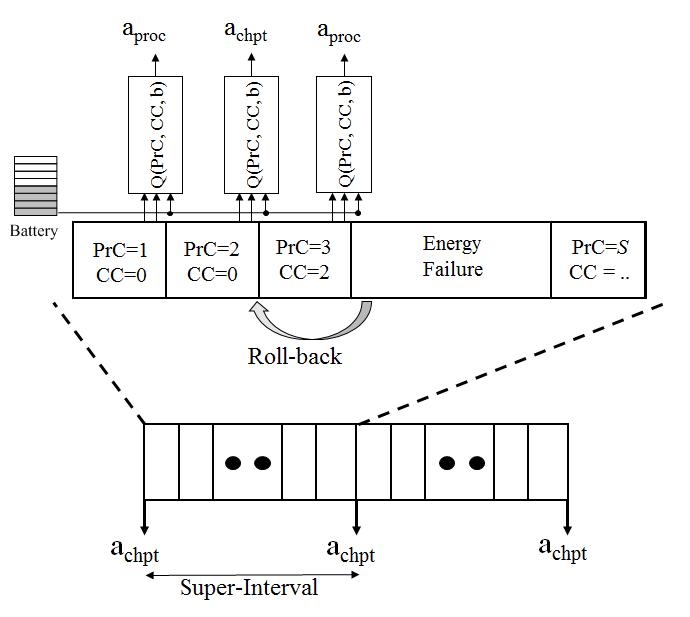}
\caption{Illustration of the checkpointing policy. A mandatory checkpoint is taken after every super interval ($S$ intervals). 
Within an interval, the checkpointing decisions are made based on the learned Q-table.}
\label{fig:cprb}
\end{figure}

\subsection{Online Decision Making}
After the Q table is learned, it is used online to derive the optimal policy for any state by picking the
action with smallest Q-value. Therefore, it is sufficient to store one \emph{action bit} corresponding to the optimum action for each state. For a small state-space, this approach entails small memory overhead and has the benefit of low energy consumption since it only requires one read from memory at the end of each interval.

The state-space of the Markov decision process grows quadratically with 
the number of intervals and hence with the dynamic execution length of the program. 
This can result in large offline training times for the Q-learning to converge, as well as large storage requirements for action bit values. We use a hybrid policy as shown in Fig~\ref{fig:cprb} that checkpoints every $S$ intervals (the super-interval). The optimal checkpointing locations \emph{within} a super-interval are determined by the proposed Q-learning approach. 
The hybrid approach limits the size of the Markov decision process state-space to $S^{2}B$ 
states, where $B$ is the number of possible battery levels. As a result, we are able to limit not only the training time, but the maximum amount of space required to store the Q-table in non-volatile main memory.

\section{Experimental Results}\label{experiments}

\subsection{System Setup}
For experimental evaluation, we used the TigerMIPS processor, 
a 32-bit 5-stage implementation of the MIPS ISA which includes 8KB L1 
instruction and data caches~\cite{tiger}.
The light-weight PRINCE block cipher 
is used to encrypt and decrypt data written to and read from the non-volatile main memory.
In order to reduce the overhead of memory writes, a TigerMIPS' write-through cache is 
replaced with a write-back cache.
An RRAM based non-volatile main memory is assumed, and NVSim~\cite{nvsim} is used 
to derive power and performance parameters for an 8 MB non-volatile memory.

The TigerMIPS RTL was modified to incorporate checkpointing and roll-back operations. 
On a checkpoint, the fetch stage is stalled until all instructions in the pipeline retire. 
The contents of PC, RF and dirty cache lines are encrypted and written back to main memory. 
The effect of energy failure is simulated by flushing the processor's pipeline, discarding the data in the instruction and data caches, 
and resetting the RF and PC. 
When energy is restored, the checkpointed state is read from main memory, decrypted and restored in the processor. 
The modified TigerMIPS processor and
PRINCE block cipher are synthesized using Cadence RTL Compiler~\cite{rc} with a 45 $nm$ technology library and 
a target 100 KHz clock. 
The $V_{dd}$ was set at $1.1V$.

\begin{figure}
\centering
\subfloat[]{%
  \includegraphics[width=0.5\textwidth]{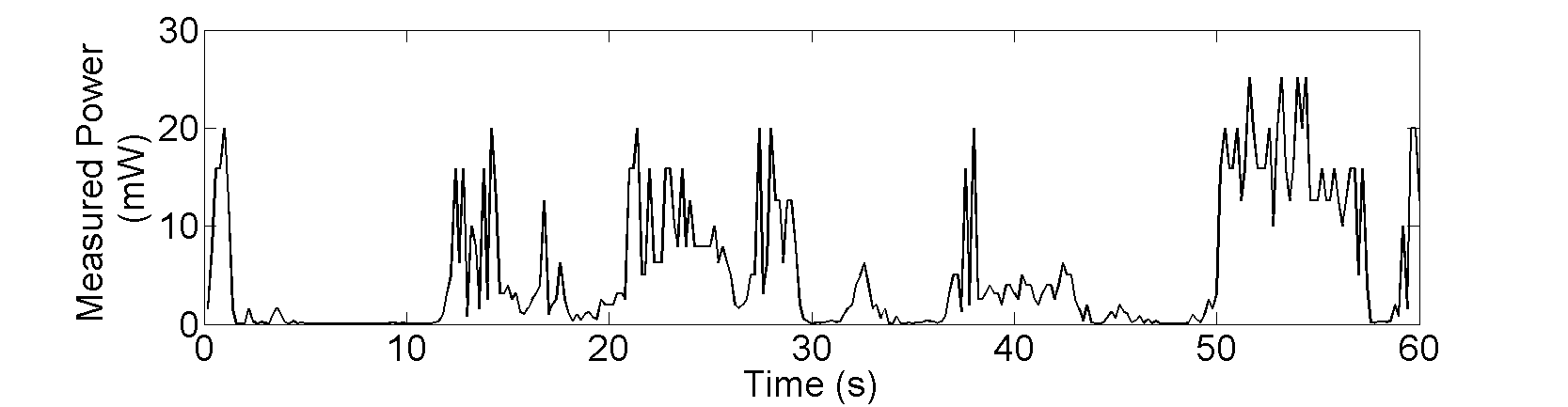}%
  \label{fig:synpa}
  }\par
  \subfloat[]{%
  \includegraphics[width=0.5\textwidth]{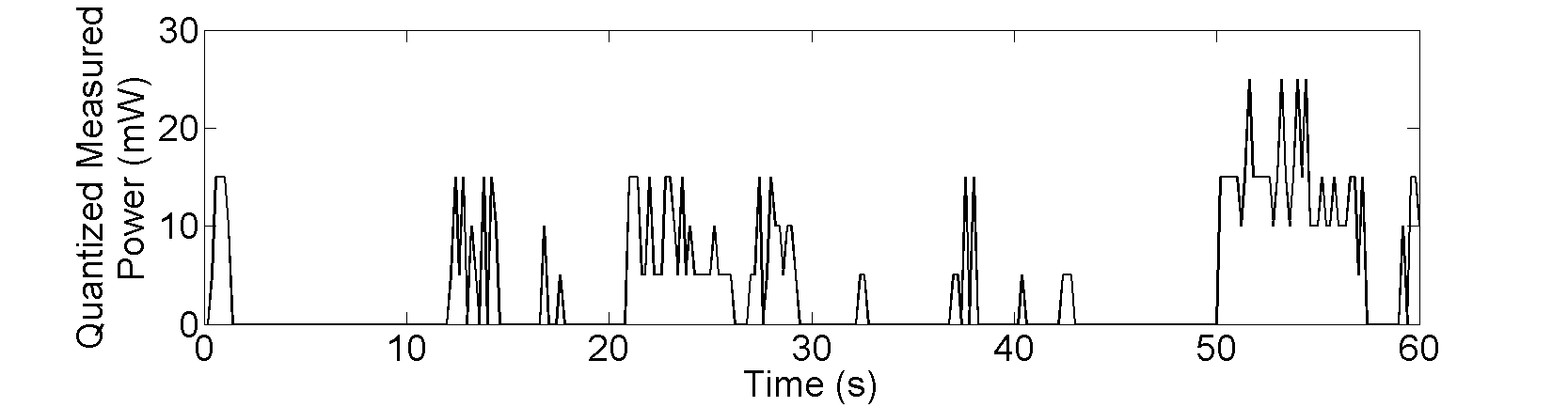}%
  \label{fig:synpb}
  }\par
\subfloat[]{%
  \includegraphics[width=0.5\textwidth]{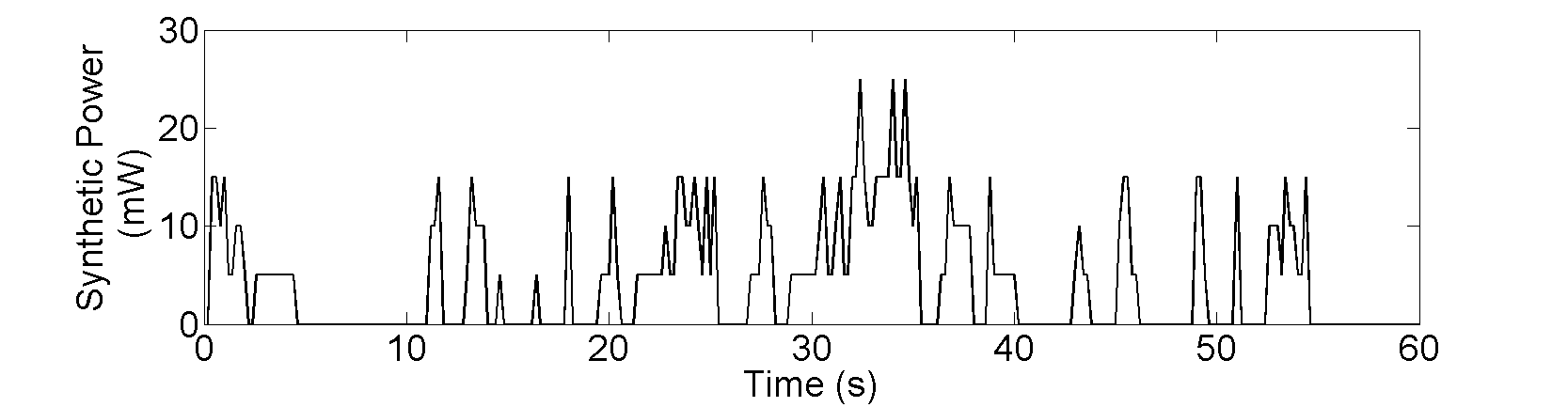}%
  \label{fig:synpc}
  }
\caption{Generating synthetic power traces based on measured harvested power from radio frequency. (a) measured radio frequency power trace from~\cite{ma:hpca}, (b) quantized measured power to six power levels, and (c) the synthetic power trace with transition probabilities extracted from the quantized measured trace.}
\label{fig:synp}
\end{figure}

For Q-learning we set the discount factor $\gamma = 1$, and $\epsilon$ (i.e., the rate at which 
random actions are picked instead of greedy ones) was reduced from $0.9$ to $0.1$ so as to perform more action exploration at the beginning. 
The size of an interval $N$ is set to 500 instructions, and a super-interval $S$ 
contains 100 intervals (i.e., a super-interval has 50K instructions).
Thus, the Progress Counter and Checkpoint Counter are small (9 bits each). 

For a super interval containing 100 intervals and 20 battery levels, the state-space has $100\times 100 \times 20 = 200K$ states. For each state, one action bit needs to be stored in memory, indicating whether a checkpoint has to be placed or not. Therefore, a memory size of $25KB$ would be sufficient to store the Q-table information. At run time, the action bit corresponding to the current state is read and the appropriate action is performed by the scheduler.

\begin{bf}{Energy Penalty Analysis}\end{bf}:
The energy penalty of storing the action bits in NVM and reading  is one read from the non-volatile memory every interval which adds up to $65.16pJ\times 100 = 6.5nJ$ per super-interval. This is only $0.002\%$ of the energy consumed by the processor in a super-interval.

We model a radio frequency based energy harvesting source as described by~\cite{ma:hpca}. Unfortunately, the sample power traces 
provided are small, while we need large traces both for training and validation. 
Modeling the harvested power as a Markov chain is common in literature~\cite{ho2010markovian}.
Thus, we
developed
a first-order Markov chain based model for the harvested 
power that transitions 
between six discrete power levels. To obtain the transition probabilities, we quantized the
measured trace to six power levels, $L=\{0, 5, 10, 15, 20, 25\}$ (power levels in $mW$).  We construct a transition matrix $T$, where $T(i,j)$ indicates the number of power level transitions from $p_i$ to $p_j$ within the trace, where $p_i, p_j \in L$. Each row $i$ of the transition matrix keeps track of the number of transitions from power level $p_i$ to all other power levels. Therefore, we can estimate the transition probabilities between power levels from the transition matrix. Fig~\ref{fig:synpa} shows the measured power from a radio frequency source. The quantized power over the six levels is shown in Fig~\ref{fig:synpb}. Once the transition matrix and the transition probability matrix is obtained, we can use the Markov chain model to obtain new synthetic power traces. An example of a synthetic power trace generated using this model is shown in Fig~\ref{fig:synpc}.

\begin{figure}
\centering
\includegraphics[width=0.4\textwidth]{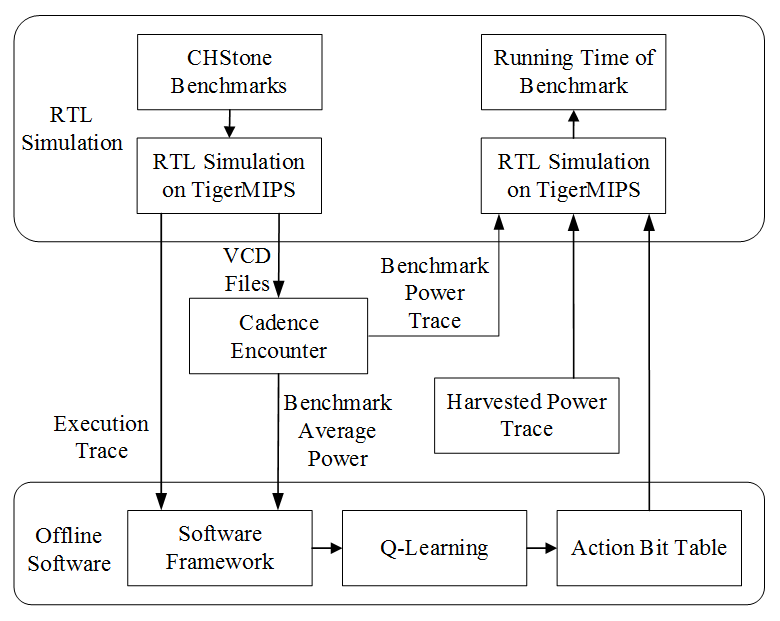}
\caption{In-house toolflow developed for experimental validation of the proposed approach.}
\label{fig:toolflow}
\end{figure}

We assume a battery with a 2 $\mu J$ capacity which is smaller than 
the $\approx 20 \mu J$ battery assumed in 
QuickRecall \cite{quick} but larger than 
the $\approx 0.5 \mu J$ capacity in \cite{ma:hpca}.

Table~\ref{tb:pars} shows the estimated 
energy consumption
for 
different components of our system, the average incoming energy, and battery parameters.

\begin{table}[h!]
\centering
\resizebox{0.45\textwidth}{!} {
    \begin{tabular} { | l | l | l | l |}
    \hline
    \bf{Parameter} & \bf{Value} & \bf{Parameter} & \bf{Value} \\ \hline \hline
    Processor & $6.3$ nJ/inst & Action Bit Memory & 25KB\\ \hline
    PRINCE  & $1.6$ nJ/8B block & Harvested (Avg) &  6 nJ/clock \\ \hline
    NV Read & $65.16$ pJ/4B line & Battery Capacity & 2 $\mu$J \\ \hline
    NV Write & $71.78$ pJ/4B line & Battery Levels & $20$\\ \hline
    \end{tabular}
}
\caption{Experimental parameters. The processor and PRINCE cipher energy consumptions 
    are obtained from RTL synthesis. The data for non-volatile memory is from NVSim~\cite{nvsim}.}
    \label{tb:pars}
\end{table}

\subsection{Tool Flow}
Our validation tool flow is shown 
in Fig~\ref{fig:toolflow} and is based on two components: an offline component that 
learns the optimal policy 
and obtains the action bit table based on the learned Q-table in \MATLAB, 
and an on-line component that uses 
detailed RTL simulations to measure execution time using our proposed and other state-of-the-art checkpointing
policies.

The offline Q-learning phase takes as input the energy harvesting traces and average of 
processor power consumption for each benchmark.
Once the optimal policy is
learned offline, the corresponding action bit table is fed 
to the RTL simulator along with dynamic traces of processor power consumption to estimate 
the program execution time with the learned policy. The power traces are generated by feeding the value change dump (VCD) files from each 
benchmark's execution to Cadence RTL Compiler.
The power traces used for
offline training and online validation are different, obtained from the same Markov chain model. Our results are presented on several benchmarks (CHStone~\cite{hara2009proposal} and FFT~\cite{legup}).

\begin{figure}
\centering
\subfloat[]{\includegraphics[width=0.25\textwidth]{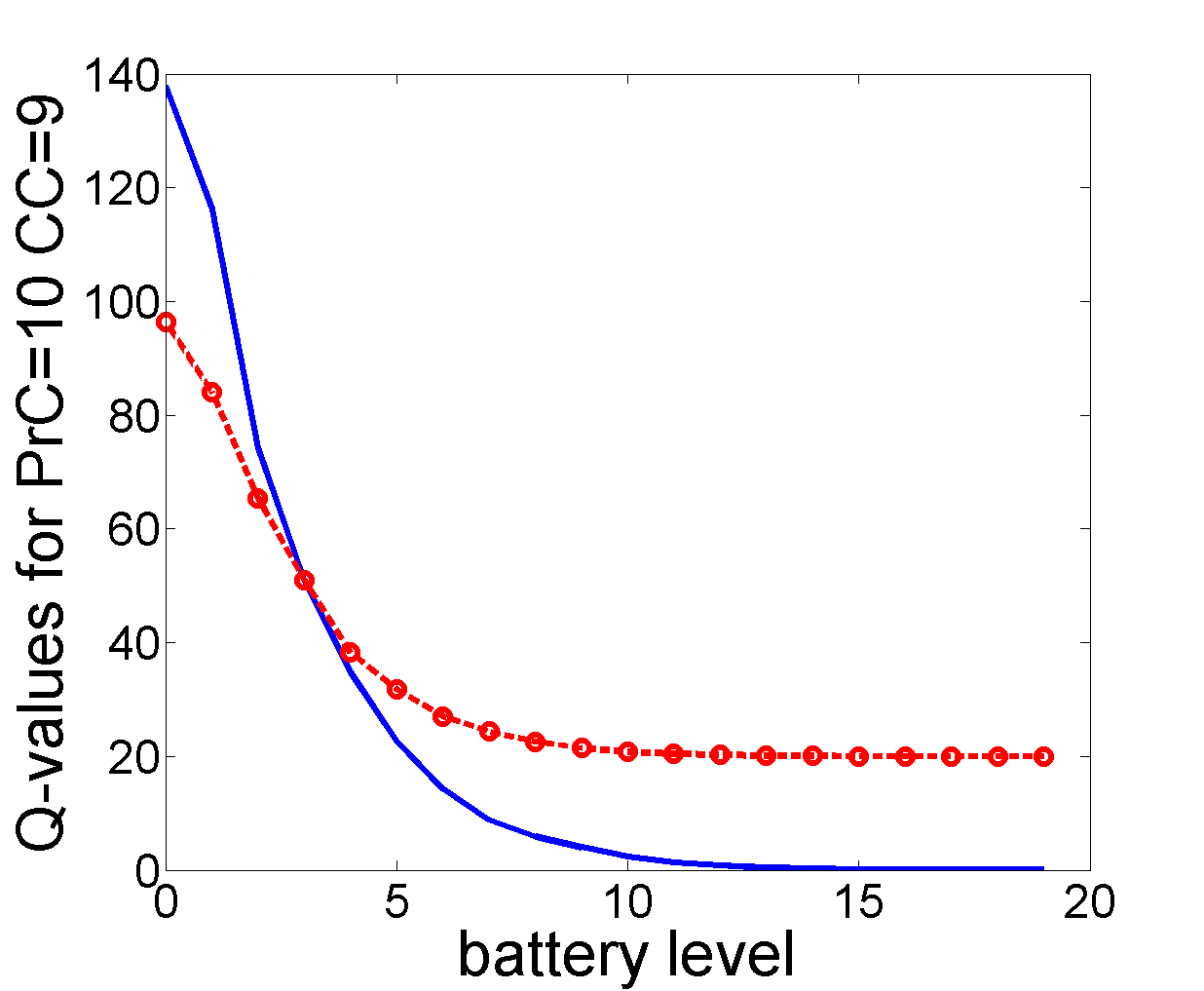}\label{fig:qvalue1}}
\subfloat[]{\includegraphics[width=0.25\textwidth]{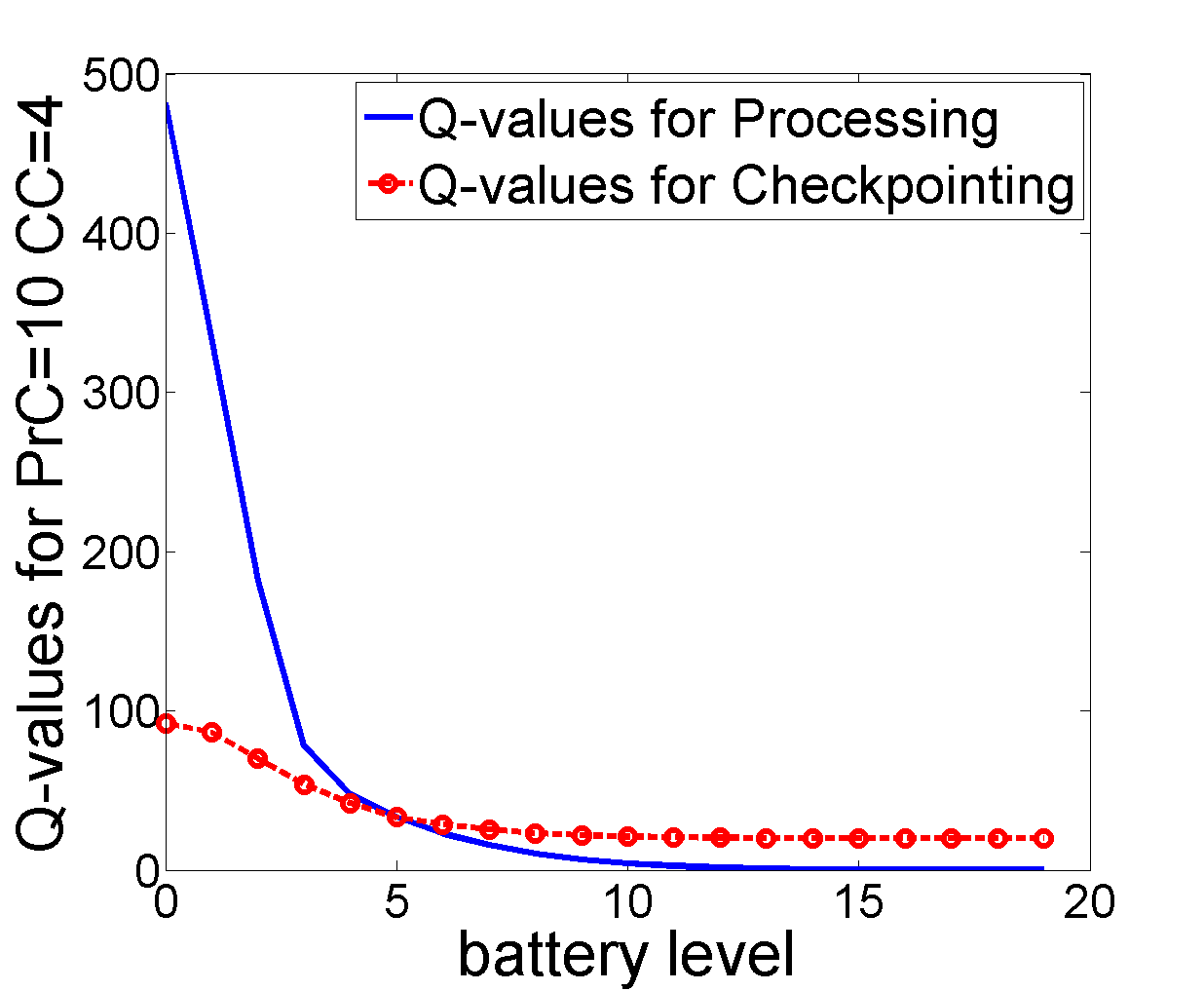}\label{fig:qvalue2}}
\caption{Converged Q-values for $a_{chpt}$ and $a_{proc}$ as a function of battery level for  
(a) $PrC=10$ and $CC=9$, i.e., at the $10^{th}$ interval with the most recent checkpoint at the $9^{th}$ interval, and 
(b) $PrC=10$ and $CC=4$.}
\label{fig:qvalue}
\end{figure}

\subsection{Comparison with Prior Art}

The Q-learning based approach is compared with 
two techniques from prior art.
The \emph{conservative} policy is based on the work in 
\cite{hibernus,ma:hpca}, where the checkpointing decisions are made \emph{only} on the current 
battery level. 
The conservative policy checkpoints 
any time the energy level in the battery 
falls below a threshold, that is equal to the energy required 
to checkpoint the PC, RF contents and the data cache. 
The processor is then turned off and  restored only when the 
battery level exceeds another threshold, which 
is at least the amount of energy required to 
read and decrypt data from the nonvolatile memory plus the energy to perform a checkpoint. 
The policy is conservative in that it
accumulates the energy required to perform a full checkpoint in battery before starting the program execution, guaranteeing that it \emph{never} incurs any roll-backs.

The second technique is a \emph{periodic}
checkpointing policy~\cite{liu2016lightweight}. A running instruction counter is maintained, and 
a checkpoint is performed any time the counter exceeds a threshold. For fairness, we 
pick the best threshold value averaged over all benchmarks (1000 instructions).


The learned Q-values as a function of battery level are shown in Fig~\ref{fig:qvalue} for checkpoint ($a_{chpt}$) and proceed with no checkpoint ($a_{proc}$) actions. In Fig~\ref{fig:qvalue1} and Fig~\ref{fig:qvalue2}, the previous checkpoint was taken one and six intervals before the current interval, respectively. 

The Q-values represent expected cost of the action, 
thus the optimal policy chooses the action with lower Q-value. We make two observations from this figure: 
(i) in both cases, 
as the battery level reduces, the checkpointing action becomes preferred to the no checkpointing action; 
(ii) however, the threshold below which the checkpointing is preferred is lower when the previous checkpoint 
was taken recently. 
In other words, policies based on a \emph{static} battery level threshold~\cite{ hibernus,ma:hpca} 
are sub-optimal.

\begin{figure}
\centering
\includegraphics[width=0.45\textwidth]{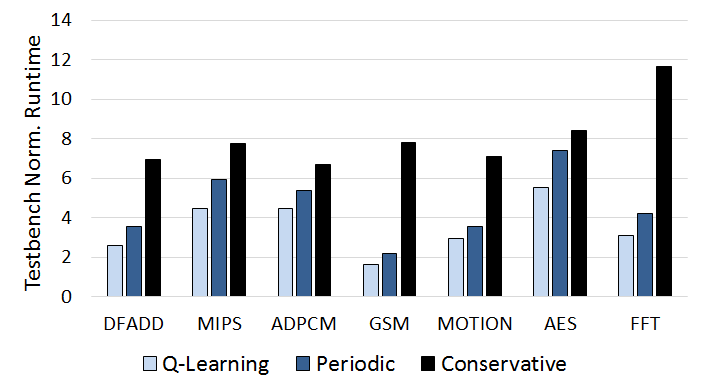}
\caption{Comparison of normalized execution times: Q-Learning based checkpointing policy 
vs periodic~\cite{liu2016lightweight} and conservative policies~\cite{hibernus,ma:hpca}. For each testbench, runtime is normalized with respect to the baseline execution time (without energy failure) for that testbench.}
\label{fig:results}
\end{figure}

\begin{table}[h!]
\centering
\resizebox{0.45\textwidth}{!} {
    \begin{tabular} { | l | c | c | c | c |}
    \hline
    \multirow{3}{*}{\bf{Benchmark}}& \multicolumn{2}{|c|}{\bf{Periodic}} & \multicolumn{2}{|c|}{\bf{Q-learning}}\\\cline{2-5}
   
    & \multirow{2}{*}{\bf{\#CPs}} & \bf{Total RB} 
    & \multirow{2}{*}{\bf{\#CPs}} &  \bf{Total RB}  \\ 
    & & \bf{Cost (s)} && \bf{Cost (s)} \\ \hline

    DFADD & 12 & 0.302 & 7 & 0.290   \\ \hline
    MIPS & 69 & 0.767 & 38 & 0.632   \\ \hline
    ADPCM & 237 & 6.13 & 139 & 5.36   \\ \hline
    GSM & 26 & 0.629 & 20 &  0.462  \\ \hline
    MOTION & 58 & 0.674 & 42 &  0.769  \\ \hline
    AES & 69 & 2.04 & 48 &  2.01  \\ \hline
    FFT & 17 & 0.391 & 11 &  0.250  \\ \hline
    \end{tabular}
}
\caption{Comparison of Q-learning based checkpointing with periodic checkpointing in terms of the number of checkpoints (CP) and roll-back (RB) cost corresponding to re-execution for various benchmarks.}
    \label{tb:statistics}
\end{table}

The normalized execution time of the 
proposed Q-learning based dynamic checkpointing policy to the periodic and conservative policies are compared in Fig~\ref{fig:results}. Normalization is done with respect to the baseline execution time with no energy failure. 
For all the benchmarks, Q-learning results in the lowest execution time followed by periodic and finally 
conservative. Specifically, Q-learning is on average $1.3\times$ faster than periodic checkpointing. 
The improvements are even greater when compared against the conservative policy. 

From Fig~\ref{fig:results} we can see that the conservative policy performs poorly compared to the other methods. Although in this method no overhead is incurred from roll-backs (and consequently lost computation), significant amount of time is lost
waiting for the battery to recharge. On the other hand, the periodic policy does not \emph{adaptively} determine 
when to checkpoint. Table~\ref{tb:statistics} shows the number of checkpoints placed by the periodic and Q-learning policies during runtime. We can see that the Q-learning based policy places fewer checkpoints compared to the periodic policy, thus incurring smaller overhead due to checkpointing. The other observation is that the re-execution cost due to roll-backs is smaller for the Q-learning policy, proving that the checkpoints are placed close to energy failure points, thus reducing the re-execution cost.

\begin{figure}
\centering
\includegraphics[width=0.4\textwidth]{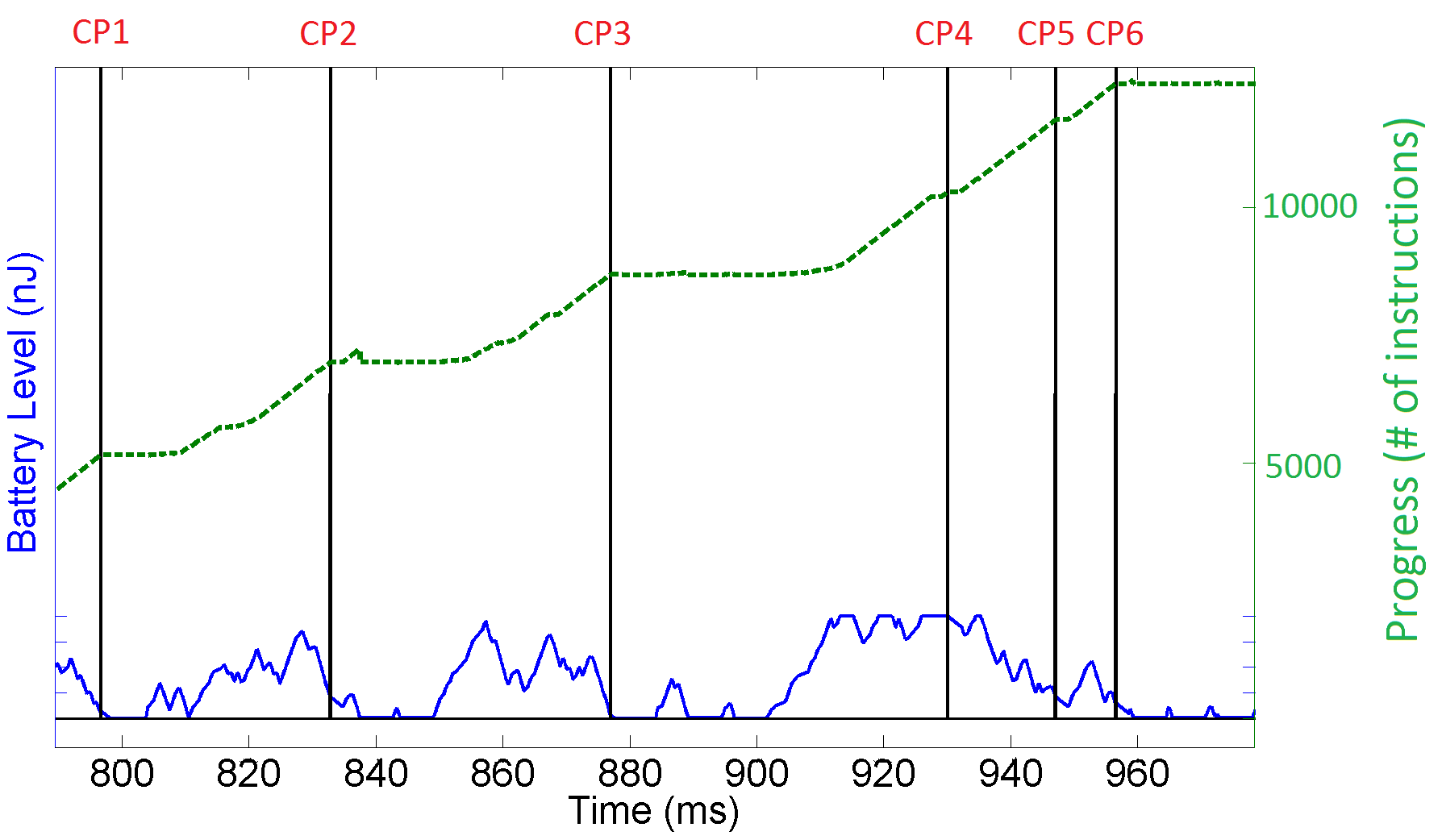}
\caption{Progress plot for the FFT benchmark using Q-learning based checkpointing. Vertical lines represent checkpoint placements. Checkpoints are triggered either by a low battery level or when a long time has passed since the previous checkpoint.}
\label{fig:progress}
\end{figure}

The execution progress of the FFT benchmark is shown in Fig~\ref{fig:progress}. This figure plots the number of (useful) instructions executed, battery level and checkpoint locations for the proposed Q-learning based checkpointing policy.
Observe that 
checkpoints can be triggered by low battery level (CP1, CP2, CP3, CP5 and CP6), or 
when a long time has 
passed since the previous checkpoint even if the battery level is high (CP4). 

\subsection{Sensitivity and Security Analysis}
We explored the effects of varying the number of possible battery levels $(B)$ and super interval size $(S)$. A simulator in \MATLAB was developed, implementing the program execution with checkpointing and roll-back. 
The simulator estimates the running time of the program by keeping track of the available energy in the battery based on the incoming harvested energy (synthetic traces) and consumed energy by the processor.

\begin{figure}
\centering
\subfloat[]{\includegraphics[width=0.17\textwidth, angle=-90]{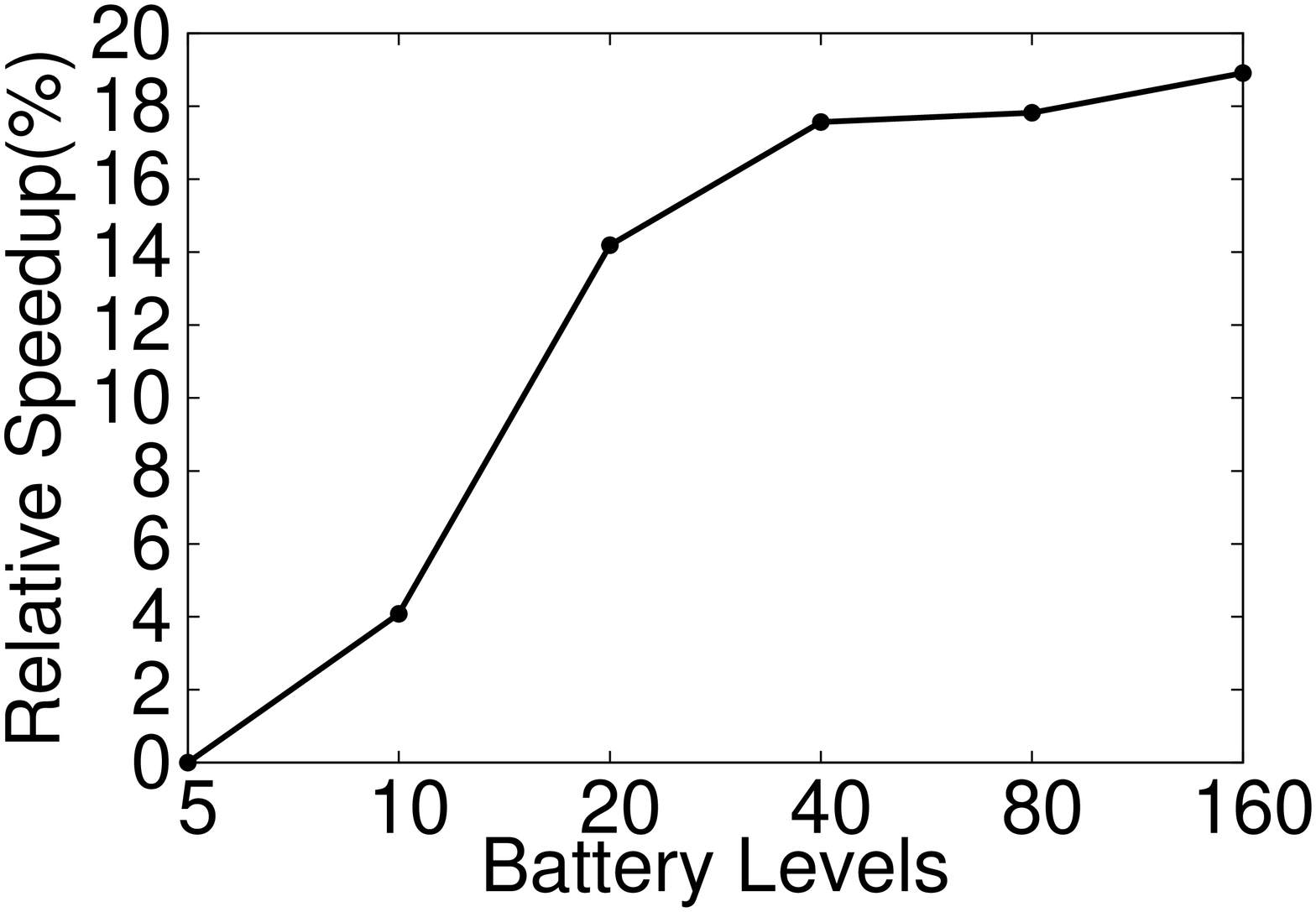}\label{fig:sens_bl}}
\subfloat[]{\includegraphics[width=0.17\textwidth, angle=-90]{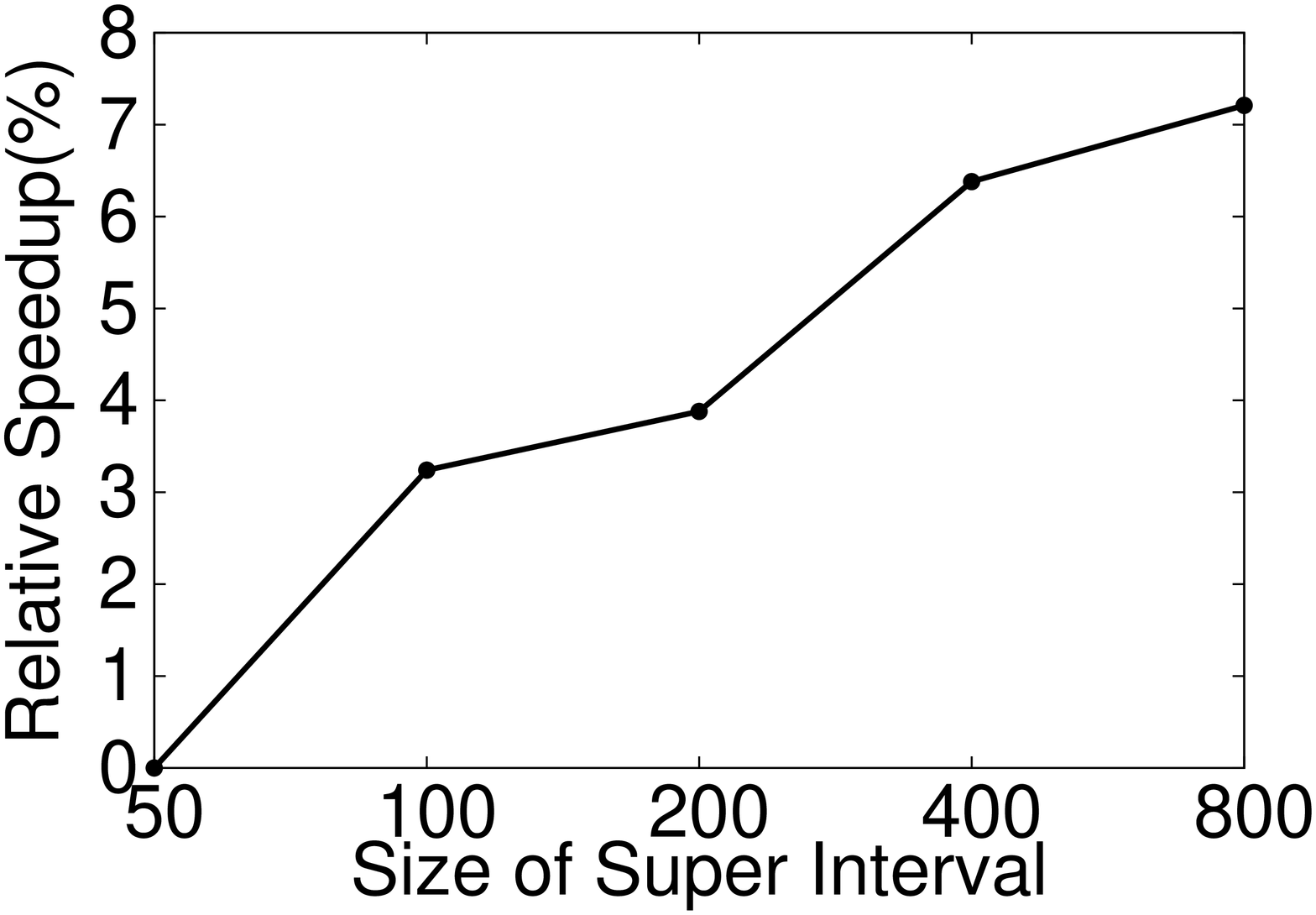}\label{fig:sens_si}}
\caption{Sensitivity analysis of the Q-learning checkpointing policy to (a) battery level B and (b) size of the super interval S. The relative speedups compared to a model with $B=5$ and $S=50$ are illustrated. With higher granularity of the battery level and a larger super interval size, the Q-learning policy performs better and program execution time decreases.}
\label{fig:sens}
\end{figure}

In the first experiment, we fixed the size of the super interval, $S=100$ and trained different models by varying $B$. The running time of the program when the Q-learning policy is used for checkpointing was estimated. Fig~\ref{fig:sens_bl} shows the relative speedup with respect to the performance of the model with $B=5$. We can see that training on higher granularity for battery level results in better performance of the Q-learning policy. Training a model with $B=160$ resulted in more than $18\%$ speedup of program execution time. However, the size of state space grows linearly with the number of battery levels, which is directly proportional to the size of memory required to store the Q-values and also results in higher training times. Note that the performance gain from increasing the battery levels begins to converge after which higher granularity would not gain much in terms of speedup.

In the second experiment, the effect of changing $S$ was examined.
We trained various models by varying $S$ and fixing $B=20$. Fig~\ref{fig:sens_si} shows the relative speedup of models based on Q-learning policy with different values of $S$ compared to the model with $S=50$. We can see that by increasing the super interval size to $800$, we get up to $7.2\%$ speedup in program execution time. Note that the size of state space and hence the memory required to store Q-values grows quadratically with the size of the super interval.

As was mentioned before, the non-volatile memory has to be encrypted at all times to prevent an attacker from reading out potentially sensitive data. However, once an attacker physically captures an IoT device, they can read out an encrypted 
image of main memory and checkpointed state. Then, by 
executing their 
own code on the device and observing the corresponding encrypted data, the attacker might 
be able to carry out a 
chosen plain-text attack to recover the on-chip encryption key. 
Although security analysis of 
PRINCE suggests that such an attack is 
impractical (requiring at least $2^{63}$ plaintexts~\cite{princean}), a cautious defender might still wish to 
to use stronger block ciphers like AES. However, AES encryption further increases the 
energy overhead of checkpointing. 
Our experiments for the FFT benchmark showed $22\%$ and $124\%$ increase in runtime when PRINCE and AES were used, respectively, for encrypting the checkpoints.
In addition to chosen plaintext attacks, the attacker might be able to launch a denial of service attack if they can tamper
with the energy source. While both the original and secure schemes are susceptible to such attacks, the energy overhead of 
secure checkpointing might increase the vulnerability to denial of service.

\section{Conclusion}\label{conclusion}
In this paper, we have proposed and evaluated a 
novel Q-learning based 
online checkpointing policy 
for secure, intermittently powered IoT devices. 
Compared to the current state-of-the-art, the proposed policy is the first 
to take into account multiple factors, including forward progress, distance 
to previous checkpoint and current battery level
to judiciously determine when to checkpoint. 
A detailed evaluation of the scheme compared to the state-of-the-art 
demonstrates up to 
$1.4\times$ reduction in execution time. We examine the effects of varying model parameters such as battery level and interval size on performance and the resulting trade-offs between model complexity and performance and memory storage requirements.
Our 
future work involves incorporating other run-time information 
in the framework, and experimenting with an energy harvester and a lightweight processor.

\bibliographystyle{unsrt}
\bibliography{main} 

\end{document}